\documentclass[12pt]{article}

\usepackage[letterpaper, margin=1in]{geometry}
\usepackage{bm}
\usepackage{graphicx}
\usepackage{caption}
\usepackage{titlesec}
\titlespacing*{\subsubsection}{0pt}{3.5ex plus 1ex minus .2ex}{0.75ex plus .2ex}
\titlespacing*{\paragraph} {0pt}{1.5ex plus 1ex minus .2ex}{1em}

\title{Numerical evidence of fluctuating stripes in the normal state of high-$T_c$ cuprate superconductors}

\author
{
\\
Edwin W. Huang,$^{1,2\ast}$ Christian B. Mendl,$^{2}$ Shenxiu Liu,$^{1,2}$\\Steve Johnston,$^{3,4}$ Hong-Chen Jiang,$^{2}$ Brian Moritz,$^{2,5}$\\Thomas P. Devereaux$^{2,6\ast}$\\
\\
\normalsize{$^{1}$Department of Physics, Stanford University, Stanford, California 94305, USA}\\
\normalsize{$^{2}$Stanford Institute for Materials and Energy Sciences, SLAC National}\\
\normalsize{Accelerator Laboratory and Stanford University, Menlo Park, CA 94025, USA}\\
\normalsize{$^{3}$Department of Physics and Astronomy, The University of Tennessee,}\\
\normalsize{Knoxville, TN 37996, USA}\\
\normalsize{$^{4}$Joint Institute for Advanced Materials, The University of Tennessee,}\\
\normalsize{Knoxville, TN 37996, USA}\\
\normalsize{$^{5}$Department of Physics and Astrophysics, University of North Dakota,}\\
\normalsize{Grand Forks, ND 58202, USA}\\
\normalsize{$^{6}$Geballe Laboratory for Advanced Materials, Stanford University,}\\
\normalsize{Stanford, CA 94305, USA}\\
\\
\normalsize{$^\ast$Correspondence to: E.W.H. (edwinwhuang@gmail.com), T.P.D. (tpd@stanford.edu)}
}

\date{}

\begin{document}

\baselineskip24pt

\maketitle

%abstract
\newpage
\begin{quote}
{\bf Upon doping, Mott insulators often exhibit symmetry breaking where charge carriers and their spins organize into patterns known as stripes. For high-$\bm{T_c}$ superconducting cuprates, stripes are widely suspected to exist in a fluctuating form. Here, we use numerically exact determinant quantum Monte Carlo calculations to demonstrate dynamical stripe correlations in the three-band Hubbard model, which represents the local electronic structure of the Cu-O plane. Our results, which are robust to varying parameters, cluster size, and boundary condition, strongly support the interpretation of a variety of experimental observations in terms of the physics of fluctuating stripes, including the hourglass magnetic dispersion and the Yamada plot of incommensurability vs.~doping. These findings provide a novel perspective on the intertwined orders emerging from the cuprates' normal state.}
\end{quote}

%main text
\newpage
Recent experiments have established charge stripes as universal in underdoped cuprate superconductors ({\it 1, 2\/}). In contrast, no consensus exists regarding the universality of spin stripes, which are present and intimately tied to charge stripes in many doped Mott insulators ({\it 1, 3--5\/}) but absent, at least in the static long-range form, in the majority of cuprates. Whether spin stripes exist in a more subtle fluctuating form in these cuprates remains an open and controversial question, of importance due to theoretical proposals suggesting a link between fluctuating stripes and the mechanism of high-Tc superconductivity ({\it 6--10\/}). The evidence for fluctuating spin stripes in the cuprates has revolved around ubiquitous observations of an hourglass-shaped magnetic excitation spectrum ({\it 11, 12\/}). Its presence both in compounds that exhibit static stripe order ({\it 13\/}) and in those that do not ({\it 14, 15\/}) finds a natural explanation in the concept of fluctuating stripes ({\it 9, 16\/}). However, alternative interpretations based on itinerant electrons exist ({\it 17\/}) and conclusive experimental evidence for fluctuating stripes remains elusive. Characterizing the nature of stripes in microscopic models provides an important alternative lens for investigating the physics of stripes in the cuprates.

Early mean-field studies of the Hubbard model ({\it 18, 19\/}) have revealed some essential attributes of stripes: a propensity for doped holes to aggregate into lines of charge that correspond to antiphase boundaries of antiferromagnetic domains. Since then, more sophisticated methods also have substantiated the presence of stripes in the ground state of the Hubbard model ({\it 20--23\/}), including recent tensor network studies indicating that superconductivity and stripes have very close ground state energies ({\it 24\/}). As these efforts have investigated only ground state properties, stripe phenomena in the disordered phase and the role of thermal fluctuations have been relatively unstudied. Furthermore, existing numerically exact, finite temperature calculations of the doped Hubbard model show only short-ranged antiferromagnetism and no sign of incommensurate spin or charge ordering ({\it 25\/}). However, these studies have been stifled by their small cluster sizes, which frustrate the antiphase behavior of stripes. Here, we overcome this obstacle with numerically exact determinant quantum Monte Carlo (DQMC) simulations on rectangular clusters substantially larger than those that have been previously considered. The horizontal dimensions are large enough to support multiple stripe domains, mitigating boundary effects that may frustrate striped correlations, while the total system size is kept sufficiently small to be computationally tractable and to avoid an unmanageable sign problem ({\it 26\/}). We also compare our results with density matrix renormalization group (DMRG) simulations on identical systems, thereby connecting zero and finite temperature results to fully characterize the presence of stripes. We choose a three-band Hubbard model of a Cu-O plane, accounting for nearest neighbor Cu-O and O-O hoppings, site energy differences, and on-site Coulomb repulsions. The DQMC simulation temperature is set to $T = 1/12\ \mathrm{eV} \approx 970\ \mathrm{K}$. Further details are provided in the Methods and Supplementary Materials, including an exploration of a range of parameters consistent with those found in the literature that yield good agreement with experiments ({\it 27\/}).

We begin by studying the $16 \times 4$ rectangular cluster with fully periodic boundary conditions. Figure~1A presents the real space, equal time spin correlation function from our finite temperature DQMC simulations at half-filling. In the undoped state, as in prior studies, copper spin correlations are dominated by commensurate antiferromagnetism, evident through the checkerboard pattern in the spin correlation function or equivalently the uniform phase of the staggered spin correlations. At $p = 0.042$ hole doping (Fig.~1B), where the doped holes predominantly reside on oxygen orbitals, antiferromagnetism persists but with decreased correlation length. Further doping reveals copper spin correlations that do not simply decay but exhibit periodic phase inversions. This can be seen in the pattern of the staggered spin correlation functions of Fig.~1B, where regions of uniform signs are separated by distinct antiphase domain walls. The presence of such domain walls is a definitive signature of stripe ordering and their periodicity of approximately $[2p]^{-1}$ agrees with results from previous works ({\it 23\/}) and presents a direct confirmation of stripe behavior in the disordered phase. To illustrate the stripes' fluctuating nature, we perform a direct comparison between DQMC and ground state DMRG simulations with identical model parameters and cluster geometry.

For the comparison, we use a cluster with periodic boundaries in the 4 unit cell vertical direction and open boundaries in the horizontal direction to break horizontal translational symmetry and potentially pin any stripe ordering. Figure~2A shows the staggered copper spin correlation function calculated by DMRG for a hole doping of $p = 1/8$. Here, antiphase domains with periodicity similar to that in the $p = 1/8$ panel of Fig.~1B are present. By varying the reference point of the correlation function, it is clear that the locations of the phase inversions are pinned by the open boundaries, corresponding to a picture of static stripes. We note that for some reference points, the nearest domain walls are sometimes shifted by one unit cell, due to contributions from short-ranged antiferromagnetic correlations, but the pinned locations of the domain walls are immediately clear by comparing with the panels for the other reference points.

This behavior stands in sharp contrast to the results from our finite temperature DQMC simulations with the same open boundary conditions and model parameters (Fig.~2B). In every panel of Fig.~2B, the structure and periodicity of the domain walls relative to the reference point are nearly identical to what is seen in the periodic boundary result of Fig.~1B. This qualitative departure from the ground state behavior seen in the DMRG simulations demonstrates that at sufficiently high temperature, stripes are delocalized and fluctuating rather than pinned by the open boundaries. For the temperature of the DQMC simulation, a lack of static long-range stripes as in the DMRG results is not surprising. Seeing vestigial signatures of the ordered phase in Fig.~1B and Fig.~2B is far less expected and provides compelling evidence for the fundamental nature of fluctuating stripes.

The elevated temperatures where stripe correlations are seen imply surprisingly strong stripe correlations over a substantial doping range. As shown in the supplement, stripe order is robust to different choices of Hubbard model parameters (Fig.~S1). Moreover, stripe order persists for larger rectangular clusters ($16 \times 6$, Fig.~S2), and additional stripes begin to develop as the transverse dimension increases ($8 \times 8$ and $10 \times 10$, Fig.~S3). This is consistent with DMRG results showing strong stripe tendencies for larger cluster sizes ({\it 20, 21\/}), indicating that our observations are not artifacts of our choice of cluster geometry. The fact that both DMRG and DQMC results mirror each other corroborates the usefulness of both methods and confirms the robustness of the measured stripe phenomena.

To draw a closer connection to experimental results, we calculate the dynamical spin structure factor $S(\bm{Q},\omega)$ by analytically continuing our DQMC data using the maximum entropy method, which is regarded as a standard procedure for extracting real-frequency spectra from imaginary-time data ({\it 28\/}). Fig.~3 displays the calculated spectra along a horizontal cut through the antiferromagnetic ordering wavevector $(\pi,\pi)$, in units where the lattice constant $a = 1$. We first consider the spectra at half-filling (Fig.~3A) as a reference. In spite of the broadening effects of the temperature ($T = 1/12\ \mathrm{eV} \approx 1/4\,J$) and finite cluster size, the structure factor exhibits a clear intensity peak and minimum in dispersion at $(\pi,\pi)$, as expected in linear spin wave theory for antiferromagnets. Upon hole doping (Fig.~3, B to F), while the high energy portions are unaffected, the soft excitations at and nearest to $(\pi,\pi)$ lose spectral weight while hardening, corresponding to the increase in the singlet-triplet gap and the destruction of antiferromagnetic ordering. In the intermediate region, at wavevectors with incommensurability corresponding to the real space periodicity in Fig.~1B, a qualitatively distinct behavior emerges. In particular, at $\bm{Q}=(3\pi/4,\pi)$ and $\bm{Q}=(5\pi/4,\pi)$ (corresponding to period-4 antiphase domain walls), systematically tracking the evolution of the structure factor with doping in Fig.~3G reveals that until roughly 1/8 hole doping, the spectral weight is maintained while the excitations soften. As the low-energy intensity peak originally at $(\pi,\pi)$ continues to bifurcate, further doping beyond 1/8 results in hardening and loss of spectral weight at $\bm{Q}=(3\pi/4,\pi)$ and $\bm{Q}=(5\pi/4,\pi)$.

This non-monotonic behavior motivates a comparison to the universal hourglass spectrum seen in inelastic neutron scattering. In Fig.~4A, we plot the center positions of MDC (momentum distribution curve) fits to our calculated structure factor for 1/8 hole doping together with experimental data from three compounds, similarly derived from MDC fits, but taken at lower temperatures (typically $10\ \mathrm{K}$). The high energy excitations at $\omega > 0.6\,J$ show a remarkable match to the neutron scattering data. For lower energies, our MDC fits do not quite resolve the neck of the hourglass; this is to be expected given the high temperature and limited momentum resolution of the DQMC simulation. The EDC fits, however, correctly resolve the collection of spectral intensity around $\omega = 0.5\,J$. With this in mind, the low energy incommensurability agrees reasonably well with the experimental results. Furthermore, the doping dependence of the incommensurability falls along the same curve as the points of the Yamada plot ({\it 12, 29\/}) (Fig.~4C). These close correspondences with well-established experimental results provide strong evidence that the three-band Hubbard model captures the microscopic features necessary to understand essential collective properties of the cuprates.

The idea that thermal and quantum fluctuations cause static stripes to melt into a fluctuating state with dynamic correlations has often been discussed theoretically ({\it 6, 9, 16\/}), but the experimental evidence remains sparse and seldom direct ({\it 30\/}). Through state-of-the-art numerical calculations, we have probed this issue in a novel manner and have shown that in the disordered phase, stripes maintain their characteristic antiphase behavior and periodicity in a fluctuating form, while being robust to variations in parameters, cluster size, and boundary condition. The fluctuating stripe order observed up to such high temperatures is a strong piece of corroborating evidence that these phenomena are strong enough to impact all electronic properties in the phase diagram. As such, the controversy between a superconducting or stripe ordered ground state in previous studies requires further clarification ({\it 24\/}). In particular, beyond comparing static properties ({\it 31\/}), a benchmark of dynamical properties determined numerically is highly desired.

%references
\newpage
\section*{References}

\begin{enumerate}
\item J.~M.~Tranquada, B.~J.~Sternlieb, J.~D.~Axe, Y.~Nakamura, S.~Uchida, {\it Nature\/} {\bf 375}, 561 (1995).
\item R.~Comin, A.~Damascelli, {\it Annu.~Rev.~Condens.~Matter Phys.\/} {\bf 7}, 369 (2016).
\item J.~M.~Tranquada, D.~J.~Buttrey, V.~Sachan, J.~E.~Lorenzo, {\it Phys.~Rev.~Lett.\/} {\bf 73}, 1003 (1994).
\item I.~A.~Zaliznyak, J.~P.~Hill, J.~M.~Tranquada, R.~Erwin, Y.~Moritomo, {\it Phys.~Rev.~Lett.\/} {\bf 85}, 4353 (2000).
\item A.~P.~Ramirez, {\it et al\/}., {\it Phys.~Rev.~Lett.\/} {\bf 76}, 3188 (1996).
\item V.~J.~Emery, S.~A.~Kivelson, O.~Zachar, {\it Phys.~Rev.~B\/} {\bf 56}, 6120 (1997).
\item S.~Kivelson, E.~Fradkin, V.~Emery, {\it Nature\/} {\bf 393}, 550 (1998).
\item J.~Zaanen, O.~Y.~Osman, H.~V.~Kruis, Z.~Nussinov, {\it Philos.~Mag.~B\/} {\bf 81}, 1485 (2001).
\item S.~A.~Kivelson, {\it et al\/}., {\it Rev.~Mod.~Phys.\/} {\bf 75}, 1201 (2003).
\item V.~Cvetkovic, Z.~Nussinov, S.~Mukhin, J.~Zaanen, {\it Europhys.~Lett.\/} {\bf 81}, 27001 (2008).
\item J.~M.~Tranquada, "Neutron Scattering Studies of Antiferromagnetic Correlations in Cuprates" in {\it Handbook of High-Temperature Superconductivity\/}, J.~R.~Schrieffer, J.~S.~Brooks, Eds. (Springer New York, 2007), chap.~6.
\item M.~Vojta, {\it Adv.~Phys.\/} {\bf 58}, 699 (2009).
\item J.~M.~Tranquada, {\it et al\/}., {\it Nature\/} {\bf 429}, 534 (2004).
\item S.~M.~Hayden, H.~A.~Mook, P.~Dai, T.~G.~Perring, F.~Doan, {\it Nature\/} {\bf 429}, 531 (2004).
\item G.~Xu, {\it et al\/}., {\it Nature Phys.\/} {\bf 5}, 642 (2009).
\item J.~Zaanen, M.~L.~Horbach, W.~van Saarloos, {\it Phys.~Rev.~B\/} {\bf 53}, 8671 (1996).
\item M.~Eschrig, {\it Adv.~Phys.\/} {\bf 55}, 47 (2006).
\item J.~Zaanen, O.~Gunnarsson, {\it Phys.~Rev.~B\/} {\bf 40}, 7391 (1989).
\item K.~Machida, {\it Physica C\/} {\bf 158}, 192 (1989).
\item S.~R.~White, D.~J.~Scalapino, {\it Phys.~Rev.~Lett.\/} {\bf 80}, 1272 (1998).
\item G.~Hager, G.~Wellein, E.~Jeckelmann, H.~Fehske, {\it Phys.~Rev.~B\/} {\bf 71}, 075108 (2005).
\item C.~C.~Chang, S.~Zhang, {\it Phys.~Rev.~Lett.\/} {\bf 104}, 116402 (2010).
\item S.~R.~White, D.~J.~Scalapino, {\it Phys.~Rev.~B\/} {\bf 92}, 205112 (2015).
\item P.~Corboz, T.~M.~Rice, M.~Troyer, {\it Phys.~Rev.~Lett.\/} {\bf 113}, 046402 (2014).
\item Y.~F.~Kung, {\it et al\/}., {\it Phys.~Rev.~B\/} {\bf 93}, 155166 (2016).
\item V.~I.~Iglovikov, E.~Khatami, R.~T.~Scalettar, {\it Phys.~Rev.~B\/} {\bf 92}, 045110 (2015).
\item C.-C.~Chen, {\it et al\/}., {\it Phys.~Rev.~Lett.\/} {\bf 105}, 177401 (2010).
\item M.~Jarrell, J.~E.~Gubernatis, {\it Phys.~Rep.\/} {\bf 269}, 133 (1996).
\item K.~Yamada, {\it et al\/}., {\it Phys.~Rev.~B\/} {\bf 57}, 6165 (1998).
\item S.~Anissimova, {\it et al\/}., {\it Nat.~Commun.\/} {\bf 5}, 3467 (2013).
\item J.~P.~F.~LeBlanc, {\it et al\/}., {\it Phys.~Rev.~X\/} {\bf 5}, 041041 (2015).
\item E.~M\"{u}ller-Hartmann, A.~Reischl, {\it Eur.~Phys.~J.~B\/} {\bf 28}, 173 (2002).
\end{enumerate}

%acknowledgements
\newpage
\section*{Acknowledgments}

We thank Arno Kampf, Steven Kivelson, Wei-Sheng Lee, Young Lee, Douglas Scalapino, Richard Scalettar, John Tranquada, and Jan Zaanen for helpful discussions. This work was supported by the U.S.~Department of Energy (DOE), Office of Basic Energy Sciences, Division of Materials Sciences and Engineering, under Contract No.~DE-AC02-76SF00515. Computational work was performed on the Sherlock cluster at Stanford University and on resources of the National Energy Research Scientific Computing Center, supported by the U.S.~DOE under Contract No.~DE-AC02-05CH11231. C.B.M.~acknowledges support from the Alexander von Humboldt Foundation via a Feodor Lynen fellowship. S.J.~is supported by the University of Tennessee's Science Alliance Joint Directed Research and Development (JDRD) program, a collaboration with Oak Ridge National Laboratory.

%main figs
\renewcommand\figurename{Fig.}

\newpage
\begin{figure}
    \centering
    \includegraphics{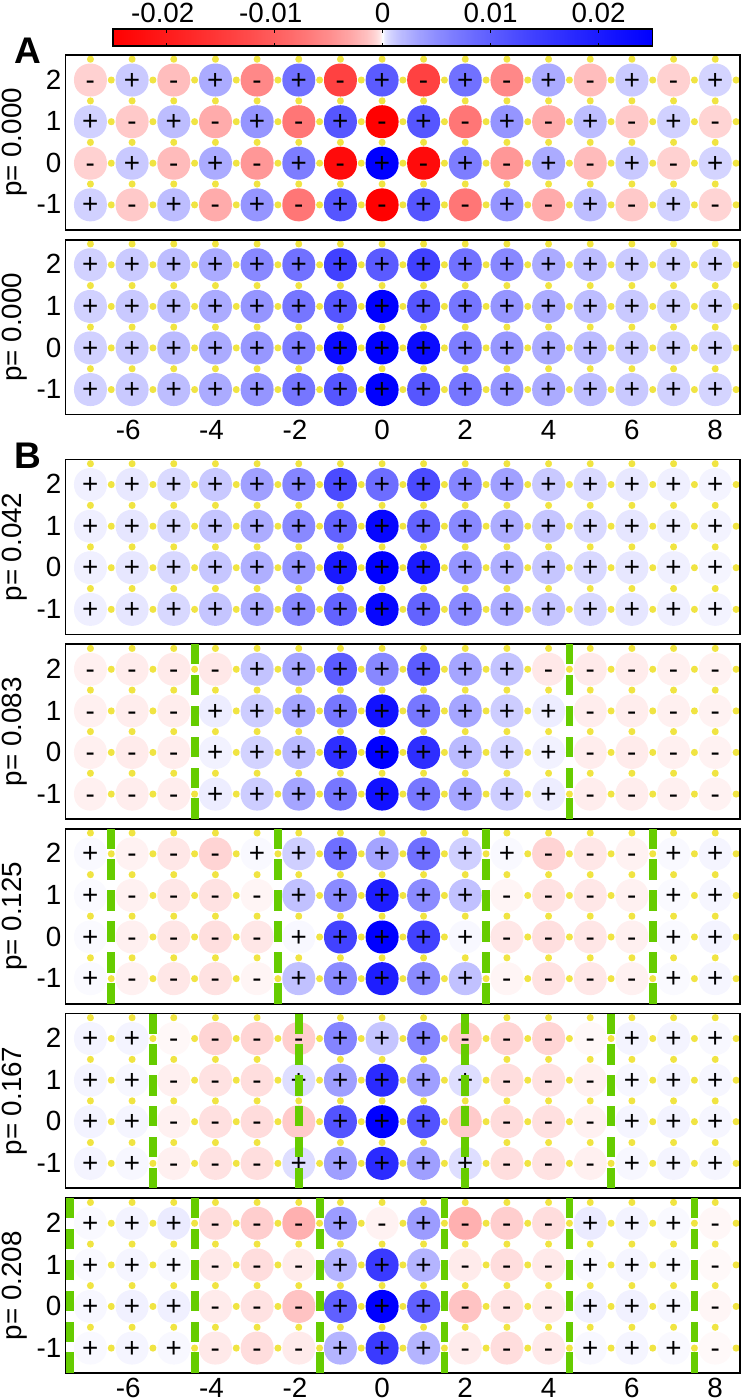}
    \caption{{\bf Stripes in the three-band Hubbard model at $\bm{T = 1/12\ \mathrm{eV}}$.} ({\bf A}) Top: Spin correlation function $S(\bm{i}-\bm{j})$. Bottom: staggered spin correlation function (inverted sign on every other lattice vector) for copper orbitals obtained by DQMC calculations on a $16 \times 4$ cluster with fully periodic boundaries at half-filling and a temperature of $T = 1/12\ \mathrm{eV}$. ({\bf B}) Staggered spin correlation functions for a range of hole doping. For clarity, the color at $(0, 0)$ is clipped. Dashed green lines indicate approximate locations of antiphase domain walls. Small yellow dots indicate positions of oxygen orbitals in the cluster. All correlations are nonzero by at least 2 standard errors (typically $4 \times 10^{-6}$).}
\end{figure}

\newpage
\begin{figure}
    \centering
    \includegraphics{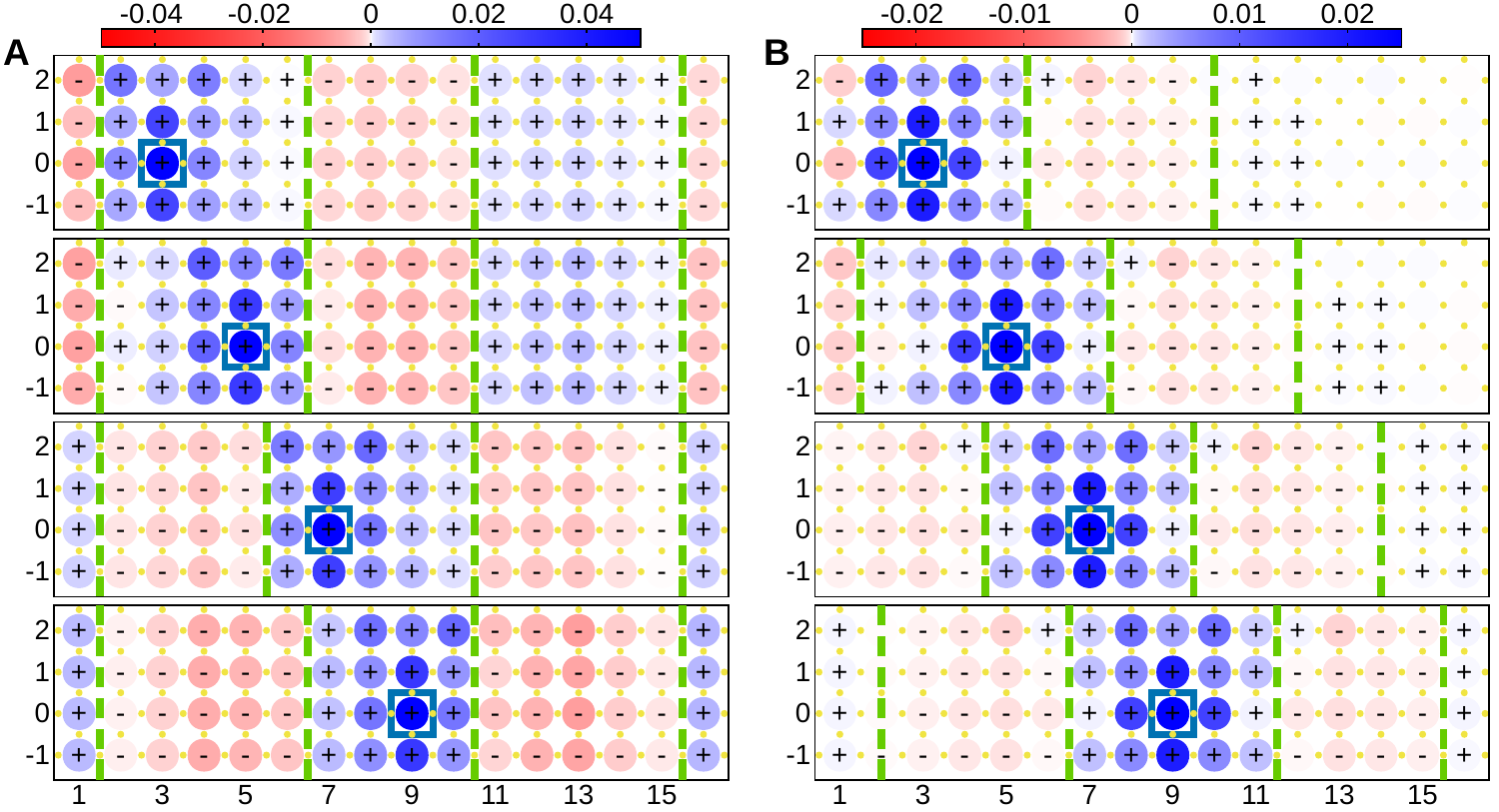}
    \caption{{\bf Comparison of static stripes ($\bm{T = 0}$) and fluctuating stripes ($\bm{T = 1/12\ \mathrm{eV}}$).} ({\bf A}) Staggered spin correlation function $S^*(\bm{i}-\bm{j})$ for copper orbitals from a DMRG simulation and ({\bf B}) a DQMC simulation. Both simulations were run with identical model parameters in a $16 \times 4$ cluster with open left and right boundaries and periodic vertical boundaries, at $p = 1/8$ hole doping. Open boundaries are terminated by oxygen orbitals, as in ({\it 25\/}). Boxes indicate reference points $\bm{i}$, which are inequivalent due to the broken translational symmetry. Colors, dashed lines, and yellow dots are as in Fig.~1. Correlations are averaged over points equivalent by symmetry. In (B), correlations showing a + or - sign are nonzero by at least 2 standard errors (typically $5 \times 10^{-6}$).}
\end{figure}

\newpage
\begin{figure}
    \centering
    \includegraphics{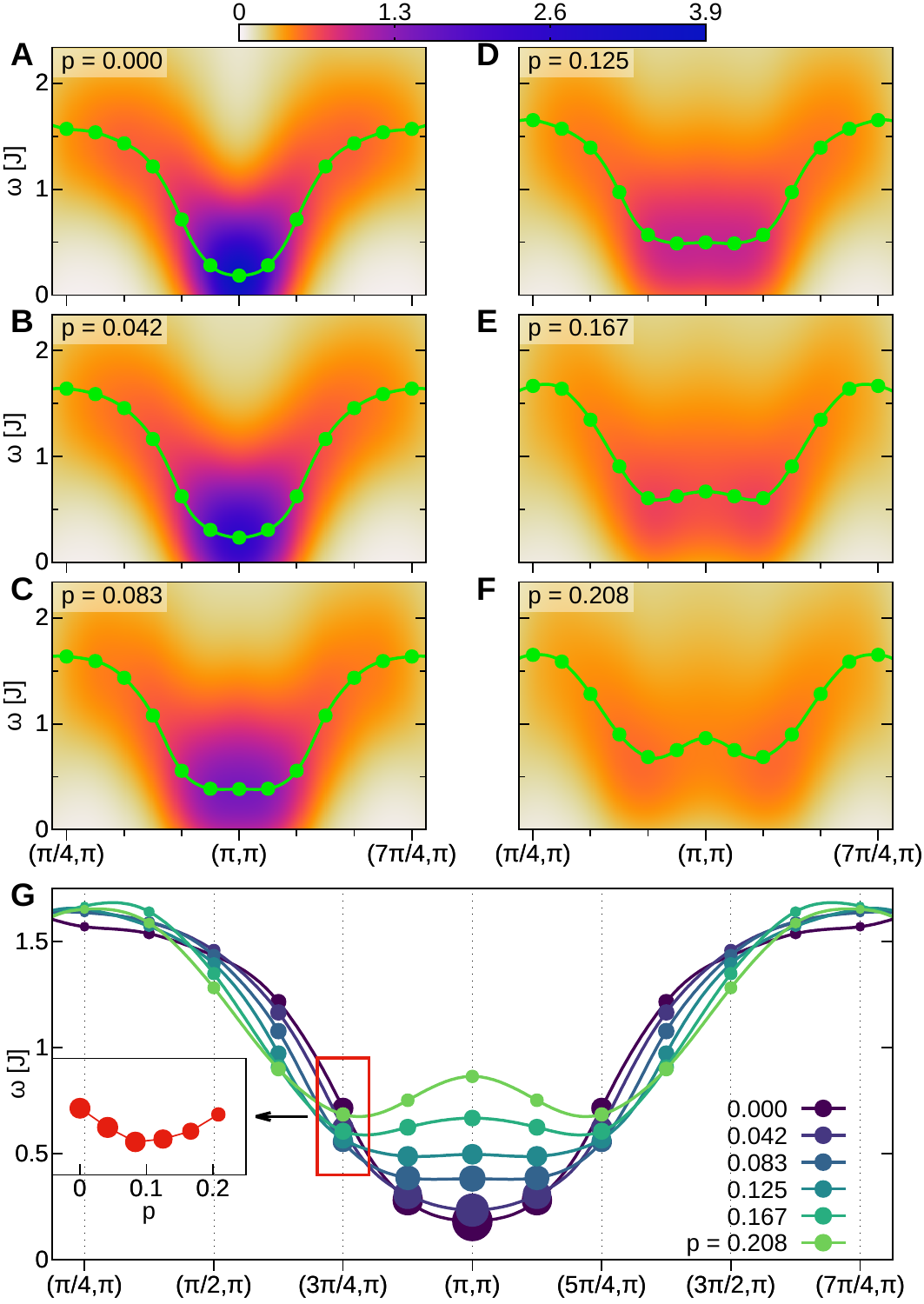}
    \caption{{\bf Magnetic excitations around $\bm{(\pi,\pi)}$.} ({\bf A} to {\bf F}) Dynamical spin structure factor $S(\bm{Q},\omega)$ along $Q_y = \pi$, calculated by MaxEnt analytic continuation ({\it 28\/}) of DQMC data at various levels of hole doping. The exchange energy ({\it 32\/}) is $J = 4 t_{pd}^{4} (U_d^{\protect\phantom{1}} + \Delta_{pd}^{\protect\phantom{1}})/(U_d^{\protect\phantom{1}} \Delta_{pd}^3) = 0.36\ \mathrm{eV}$ for our parameters. Cluster geometry is $16 \times 4$ with periodic boundaries, corresponding to a momentum resolution of $\pi/8$ in the horizontal direction. Spline interpolation is applied to approximate spectra at interlying wavevectors. Green lines represent EDC (energy distribution curves) centers, determined via Lorentzian fits, with dots at nonuninterpolated wavevectors. ({\bf G}) Superimposed EDC centers. Dot diameter represents integrated spectral weight of each EDC. Inset: EDC centers for $\bm{Q} = (3\pi/4,\pi)$ with hole doping on the x-axis.}
\end{figure}

\newpage
\begin{figure}
    \centering
    \includegraphics[width=16cm]{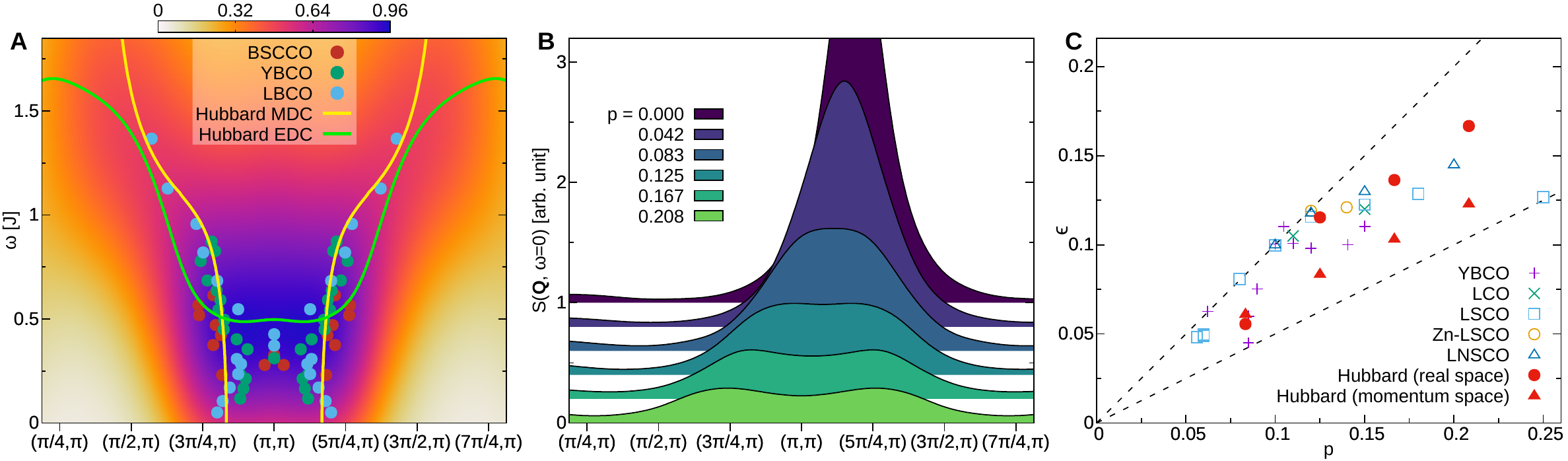}
    \caption{{\bf Comparisons against experimental results.} ({\bf A}) Dynamical spin structure factor $S(\bm{Q},\omega)$ and EDC centers along $Q_y = \pi$ for 1/8 hole doping, as in Fig.~3D. Yellow lines represent centers of double Gaussian fits to MDCs (momentum distribution curves). Dots show data from inelastic neutron scattering in ({\it 12\/}, {\it 16\/}), and references therein. The Cu-O plane hole doping is $p \approx 0.16$ for optimally doped $\mathrm{Bi_2Sr_2CaCu_2O_{8+x}}$, $p \approx 0.1$ for $\mathrm{YBa_2Cu_3O_{6.5}}$, and $p = 0.125$ for $\mathrm{La_{1.875}Ba_{0.125}CuO_4}$. ({\bf B}) Waterfall plot of MDCs for $\omega = 0$. ({\bf C}) Low-energy spin incommensurability $\epsilon$ vs. hole doping $p$. In units where $a = 1$, $\epsilon$ is the separation from $\bm{Q} = (\pi,\pi)$, divided by $2\pi$. Dashed lines show $\epsilon = p$ and $\epsilon = p/2$, corresponding to incommensurabilities of half-filled stripes ({\it 20\/}) and filled stripes ({\it 16\/}), respectively. Data from ({\it 13\/}, {\it 14\/}), and references therein are plotted with estimates of incommensurability from the DQMC data. The real space estimate is half the inverse of the antiphase domain walls periodicity in Fig. 1B. The momentum space estimate is obtained by double Gaussian fits to MDCs in (B). Neither methods show incommensurability for $p = 0.042$.}
\end{figure}

%supplement
\clearpage
\section*{\centering \LARGE Supplementary Materials}

%methods
\section*{Methods}

\paragraph{Hubbard model:}
We consider the three-band Hubbard model defined by the following Hamiltonian ({\it 25\/}, {\it 26\/}, {\it 33\/}):
$$H = \sum_{\langle i,j \rangle, \sigma} t^{pd}_{ij} \left( d^{\dagger}_{i\sigma} c^{\phantom{1}}_{j\sigma} + h.c.\right) + \sum_{\langle j,j' \rangle, \sigma} t^{pp}_{jj'} \left( c^{\dagger}_{j\sigma} c^{\phantom{1}}_{j'\sigma} + h.c.\right) - \mu \sum_{i,\sigma} n^d_{i\sigma}$$
$$\quad + (\Delta_{pd} - \mu) \sum_{j,\sigma} n^p_{j\sigma} + U_d \sum_i n^d_{i\uparrow} n^d_{i\downarrow} + U_p \sum_j n^p_{j\uparrow} n^p_{j\downarrow}.$$

The model accounts for the copper $3 d_{x^2-y^2}$ orbitals and oxygen $2p_x$ and $2p_y$ orbitals. Here, $d^{\dagger}_{i\sigma}$ and $d_{i\sigma}$ are the hole creation and annihilation operators with spin $\sigma \in \{\uparrow, \downarrow\}$ for the copper $d_{x^2-y^2}$ orbital at site $i$. Analogously, $c^{\dagger}_{j\sigma}$ and $c_{j\sigma}$ are the hole creation and annihilation operators for the oxygen $p_x$ or $p_y$ orbital at site $j$. $t_{ij}^{pd}$ and $t_{jj'}^{pp}$ describe the copper-oxygen and oxygen-oxygen hopping amplitudes, respectively, including orbital phase factors ({\it 25\/}, {\it 33\/}). $n^{d}_{i\sigma} = d^{\dagger}_{i\sigma} d^{\phantom{1}}_{i\sigma}$ and $n^{p}_{i\sigma} = c^{\dagger}_{i\sigma} c^{\phantom{1}}_{i\sigma}$ are the number operators. $\Delta_{pd}$ is the difference in site energies of the oxygen and copper orbitals. $U_d$ and $U_p$ are the on-site Coulomb interaction strengths of the respective orbitals. $\mu$ is the chemical potential used in DQMC to control the average occupancy; in DMRG, a fixed occupancy is prescribed in the canonical ensemble.

\paragraph{DQMC algorithm:}
We simulate the three-band Hubbard model with DQMC ({\it 34\/}, {\it 35\/}) using the following parameter set (in units of eV): $U_d=6.0$, $U_p=0.0$, $t_{pd}=1.13$, $t_{pp}=0.49$, and $\Delta_{pd}=3.0$. All parameters are standard ({\it 36\/}) except for the Coulomb interactions, which are slightly reduced to ameliorate the sign problem. We confirm in Fig.~S1 and the accompanying section of the Supplementary Text that the reduction has no qualitative effect on the spin response. The chemical potential is adjusted as needed to achieve the desired doping level to an accuracy of $O(10^{-4})$ or better. We choose an inverse temperature of $\beta=12\ \mathrm{eV}^{-1}$ with imaginary time steps of $0.125\ \mathrm{eV}^{-1}$. Between 500 and 5000 independently seeded Markov chains, each with around 50000 equilibration and 200000 measurement sweeps, are run for each doping level.

\paragraph{DQMC error analysis:}
The average signs of the simulations of the $16 \times 4$ clusters in the main text and the $8 \times 8$ clusters of Fig.~S3 range from $0.42$ at half-filling to $0.22$ at $0.125 \leq p \leq 0.208$. For the larger $16 \times 6$ and $10 \times 10$ simulations in Figs.~S2 and S3, the average signs of both are $0.10$. Detailed plots of the doping and temperature dependence of the average sign may be found in ({\it 25\/}, {\it 26\/}).
Observables are calculated after combining data from the independent Markov chains, with average values and sampling errors estimated through Jackknife resampling. Due to the large number of samples, standard errors in the correlation functions are less than $10^{-5}$ (Fig.~S8).

\paragraph{DMRG algorithm:}
Using the same model parameters, we perform the standard DMRG simulations ({\it 37, 38\/}) with up to 50 sweeps and keep up to $m = 8000$ DMRG block states with a typical truncation error of $6 \times 10^{-6}$ per step. This led to excellent convergence for the results that we report here.

\paragraph{Analytic continuation:}
We apply the maximum entropy method ({\it 28\/}, {\it 39\/}) to calculate dynamical structure factors from the imaginary-time correlation functions measured by DQMC. A non-informative model function is used, as described in ({\it 28\/}), and we find our final spectra to depend little on the choice of model function.

%supp text
\newpage
\section*{Supplementary Text}

\subsubsection*{Generality of results for different parameter sets and cluster geometry}
To ensure that the results presented in the main text are not unique to their parameter set (listed in Methods), we consider the effects of varying each degree of freedom in the three-band Hubbard model, within a reasonable range that is applicable to the cuprates. Figure~S1 displays the staggered spin correlation functions from DQMC simulations of $16 \times 4$ clusters for various parameter sets. It is immediately clear that stripes are always present, regardless of variations in the parameters. Additionally, the periodicity of the stripes is nearly unaffected by the changes in parameters. We note for some parameter sets, a worsened fermion sign problem requires an increase in the temperature to maintain acceptable statistics. As this decreases the correlation length, the signal-to-noise ratio for data near the boundaries is reduced. In spite of this, our data show for every parameter set, at least one point near the edge shows a positive value above two standard errors.

We also investigate a slightly larger $16 \times 6$ cluster to verify that the $16 \times 4$ geometry is not uniquely conducive to stripe formation. Figure~S2 shows the staggered spin correlation function for the larger cluster at one level of doping. The computational cost and worsened sign problem associated with the larger cluster prevents a more thorough investigation. However, it is still clear that the same periodic antiphase domain walls in the $16 \times 4$ geometry (Fig.~1) are present. Taken together, these observations provide compelling evidence that our conclusions reflect generic properties rather than ones specific to a special geometry or to a particular region in parameter space.

\subsubsection*{Square cluster geometry results}
In addition to the rectangular geometries, we also consider a more conventional $8 \times 8$ square cluster with periodic boundaries. Staggered spin correlation functions from our DQMC calculations are shown in Fig.~S3. At half-filling (Fig.~S3C), antiferromagnetism dominates as in the rectangular result (Fig.~1A). With hole doping (Fig.~S3D), the square cluster also shows phase inversions, but here it is seen only in the corners. This behavior can be understood simply by considering the symmetry of the square cluster. After averaging over the sampled configurations, the Monte Carlo does not break the rotational symmetry of the square cluster. Therefore if stripes are present, the calculation must reflect an equal superposition of horizontal and vertical stripes. Fig.~S3B depicts an idealized picture of the expected pattern in the staggered spin correlations for such a superposition of period-4 stripes. The correspondence to the data in Fig.~S3D corroborates our rectangular geometry results and shows that stripe formation is not merely an artifact of the rectangular cluster. The agreement with the larger $10 \times 10$ cluster (Fig.~S3E) proves this behavior is not unique to any specific cluster size. Additionally, the observed pattern in Fig.~S3D is incompatible with that of a superposition of diagonal stripes (Fig.~S3A). This establishes the axial nature of the stripes in the three-band Hubbard model, in accordance with experimental knowledge of the cuprates.

\subsubsection*{Charge modulations in DQMC and DMRG}
Figure~S4 plots the staggered spin correlation functions and occupancy profiles from a DMRG simulation using a set of parameters with the full interaction strength. As in Fig.~2A, the $16 \times 4$ cluster with cylindrical boundaries is employed, giving rise to pinned domain walls in the spin correlations. Examining the occupancy profile (Fig.~S4B) reveals charge modulations with peaks aligned to the antiphase domain walls. This result agrees with a previous DMRG study ({\it 25\/}) of the three-band model that considered a smaller $8 \times 4$ cluster. For the simulations in Fig.~4, which have reduced interaction strengths (see Methods), the corresponding occupancy profiles are plotted in Fig.~S5. The DMRG simulation (Fig.~S5A) again shows charge modulations, though the connection to the antiphase domain walls (Fig.~2A) is not as distinct. In contrast to these DMRG simulations, the finite temperature data of the DQMC calculations (Fig.~S4B) show no charge modulation in the bulk, with only rapid, short-ranged oscillations at the open boundaries. This difference can be traced to the unpinned domain walls seen in the spin correlations (Fig.~2B) and agrees with a picture of delocalized, fluctuating stripes.

\subsubsection*{Details of dynamical spin structure factor analysis}
The centers of the dynamical spin structure factor's EDCs in Fig.~3 and Fig.~4A are determined by a fit to a single Lorentzian peak. Examples of the fit are provided in Fig.~S6. In Fig.~4A, we also analyze the MDCs of the dynamical spin structure factor by fitting to two Gaussian peaks, symmetric about $(\pi,\pi)$, plus a uniform background. This procedure is essentially similar to the fitting of inelastic neutron scattering data used to present the hourglass dispersion. Examples of the fit are shown in Fig.~S7, B and C. Plotting the Gaussians' centers against energy and hole doping level (Fig.~S7A) provides another view of the doping dependence of the magnetic excitations. As in Fig.~3, the high energy excitations are minimally affected by hole doping. For the low energy portions of the spectra, at half-filling and at $p = 0.042$ hole doping, the broadness of the spectra results in fits to essentially a single Gaussian, up to around $\omega = 0.65 J$. The development of incommensurability is faintly visible at $p = 0.083$ and progressively increases with doping, as also visible in Fig.~4B and in agreement with the decreasing real-space periodicity of the antiphase domain walls in Fig.~1B.

%supp references
\newpage
\section*{References (Supplementary Materials)}

\begin{enumerate}
\setcounter{enumi}{32}
\item G.~Dopf, A.~Muramatsu, W.~Hanke, {\it Phys.~Rev.~Lett.\/} {\bf 68}, 353 (1992).
\item R.~Blankenbecler, D.~J.~Scalapino, R.~L.~Sugar, {\it Phys.~Rev.~D\/} {\bf 24}, 2278 (1981).
\item S.~R. White, {\it et al\/}., {\it Phys.~Rev.~B\/} {\bf 40}, 506 (1989).
\item C.-C. Chen, {\it et al\/}., {\it Phys.~Rev.~Lett.\/} {\bf 105}, 177401 (2010).
\item S.~R. White, {\it Phys.~Rev.~Lett.\/} {\bf 69}, 2863 (1992).
\item E.~Stoudenmire, S.~R.~White, {\it Annu.~Rev.~Condens.~Matter Phys.\/} {\bf 3}, 111 (2012).
\item A.~Macridin, S.~P.~Doluweera, M.~Jarrell, T.~Maier, http://arXiv.org/abs/cond-mat/0410098 (2004).
\end{enumerate}

%supp figures
\renewcommand{\thefigure}{S\arabic{figure}}
\setcounter{figure}{0}

\newpage
\begin{figure}
    \centering
    \includegraphics{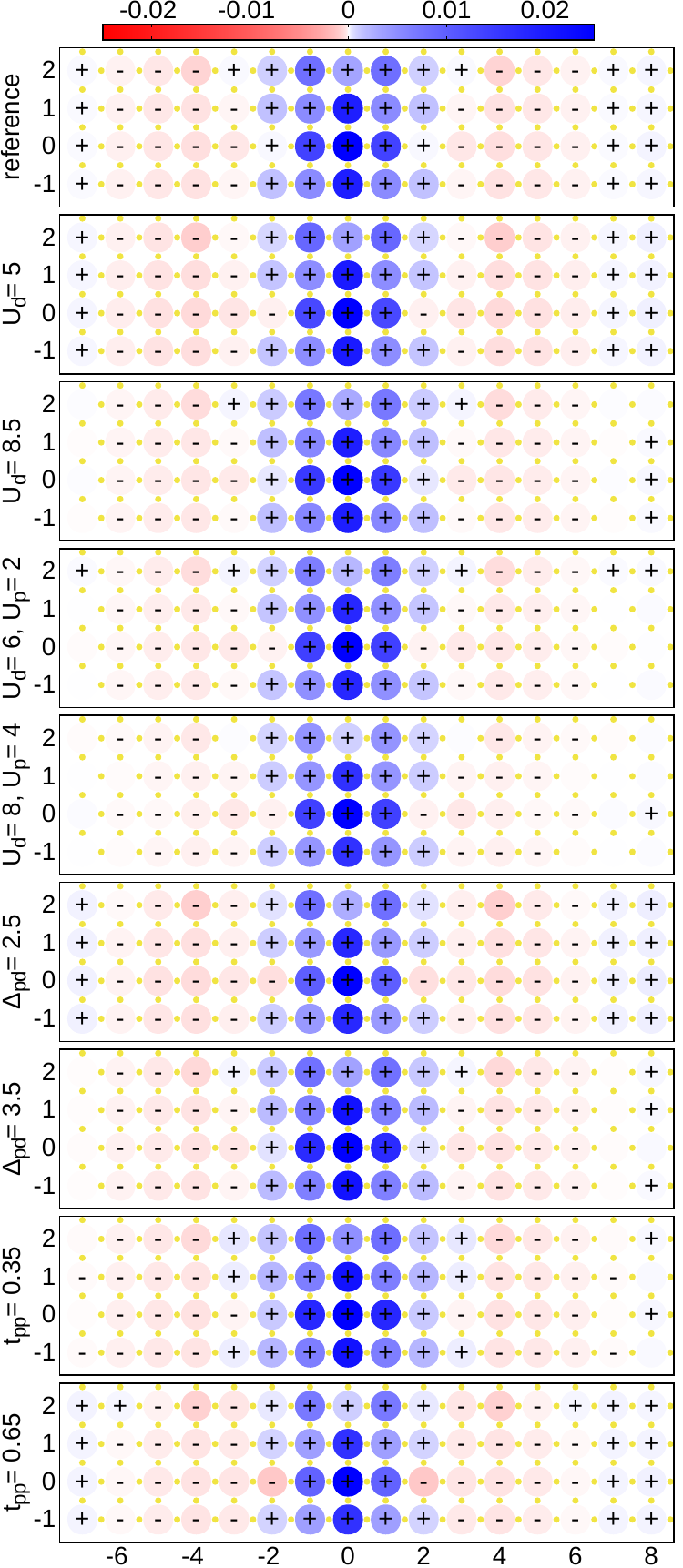}
    \caption{{\bf Effects of varying model parameters.} Staggered spin correlation function from DQMC calculations of the three-orbital Hubbard model at $p = 1/8$ hole doping. Top panel corresponds to a simulation with the parameters listed in Methods, and is identical to the center panel of Fig.~1B. Remaining panels display the varied parameters on the y-axis labels. Because changing the various parameters affects the sign problem, the simulation temperature is set as low as possible while keeping the average sign above 0.1. From top to bottom, the inverse temperatures in $\mathrm{eV}^{-1}$ are: 12.0, 14.4, 10.0, 10.0, 8.0, 15.6, 11.04, 12.0, 12.0. Correlations showing a + or - sign are nonzero by at least 2 standard errors (typically $5 \times 10^{-6}$).}
\end{figure}

\newpage
\begin{figure}
    \centering
    \includegraphics{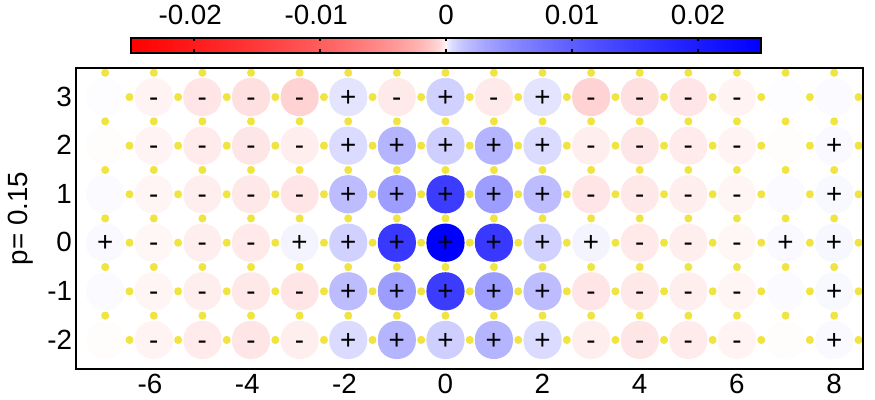}
    \caption{{\bf Stripes in a DQMC simulation of a $\bm{16 \times 6}$ cluster.} Staggered spin correlation functions at $p = 0.15$ hole doping for a $16 \times 6$ cluster with periodic boundaries. Correlations showing a + or - sign are nonzero by at least 2 standard errors (typically $3 \times 10^{-6}$).}
\end{figure}

\newpage
\begin{figure}
    \centering
    \includegraphics{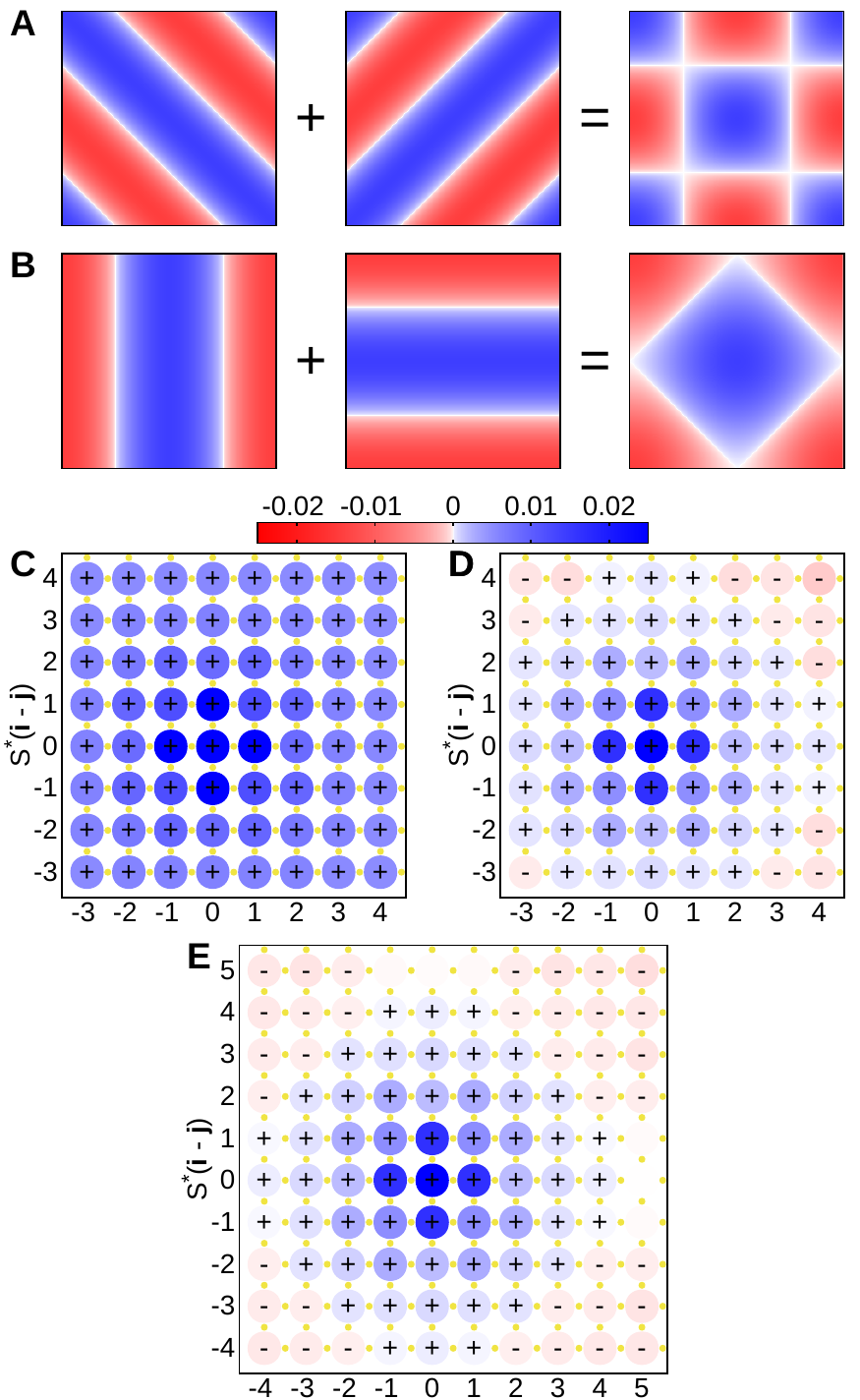}
    \caption{{\bf DQMC calculations on square clusters.} ({\bf A}) Expected pattern in staggered spin correlation function for a superposition of diagonal stripes and ({\bf B}) of axial stripes. ({\bf C}) Staggered spin correlation functions from a DQMC simulation on an $8 \times 8$ cluster at half-filling, ({\bf D}) on an $8 \times 8$ cluster at 1/8 hole doping, and ({\bf E}) on a $10 \times 10$ cluster at 1/8 hole doping. For clarity, the color scale is centered at zero and the color at $(0,0)$ is clipped. Small yellow dots indicate positions of oxygen orbitals in the cluster. In (C) and (D), all correlations are nonzero by at least 8 standard errors (typically $7 \times 10^{-6}$). In (E), correlations showing a + or - sign are nonzero by at least 2 standard errors (typically $1 \times 10^{-5}$).}
\end{figure}

\newpage
\begin{figure}
    \centering
    \includegraphics{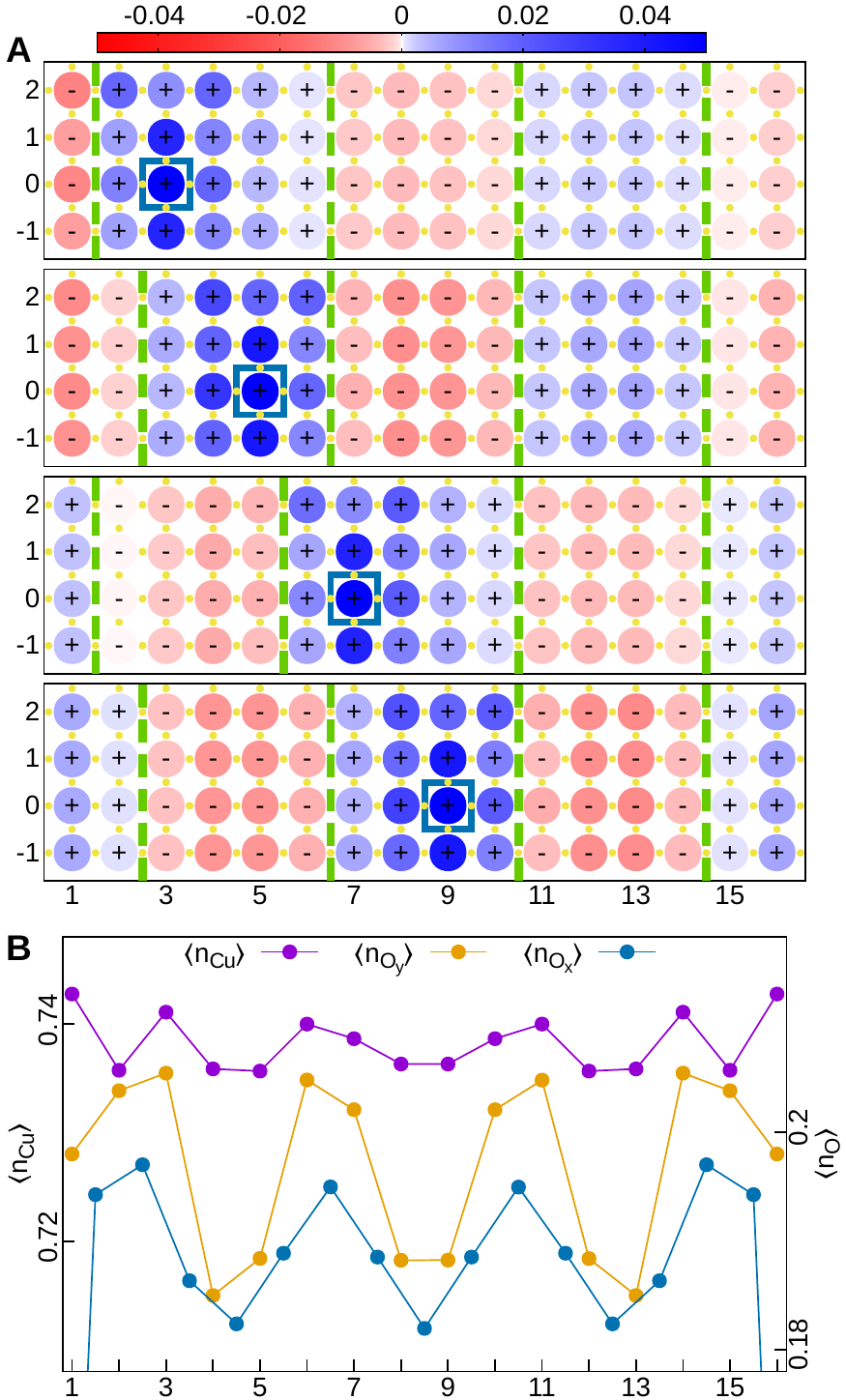}
    \caption{{\bf Static stripes in a DMRG simulation.} ({\bf A}) Staggered spin correlation functions from a DMRG simulation with the following set of parameters: $p = 0.125$, $U_d = 8.5$, $U_p = 4.1$, $t_{pd} = 1.0$, $t_{pp} = 0.49$, and $\Delta_{pd} = 3.24$. Cluster geometry is identical to that in Fig.~2 of the main text: $16 \times 4$ with open left and right boundaries and periodic vertical boundaries. The boxes, colors, dashed lines, and yellow dots are also as in Fig.~2. ({\bf B}) The same DMRG simulation's occupancies for each orbital as a function of horizontal position.}
\end{figure}

\newpage
\begin{figure}
    \centering
    \includegraphics{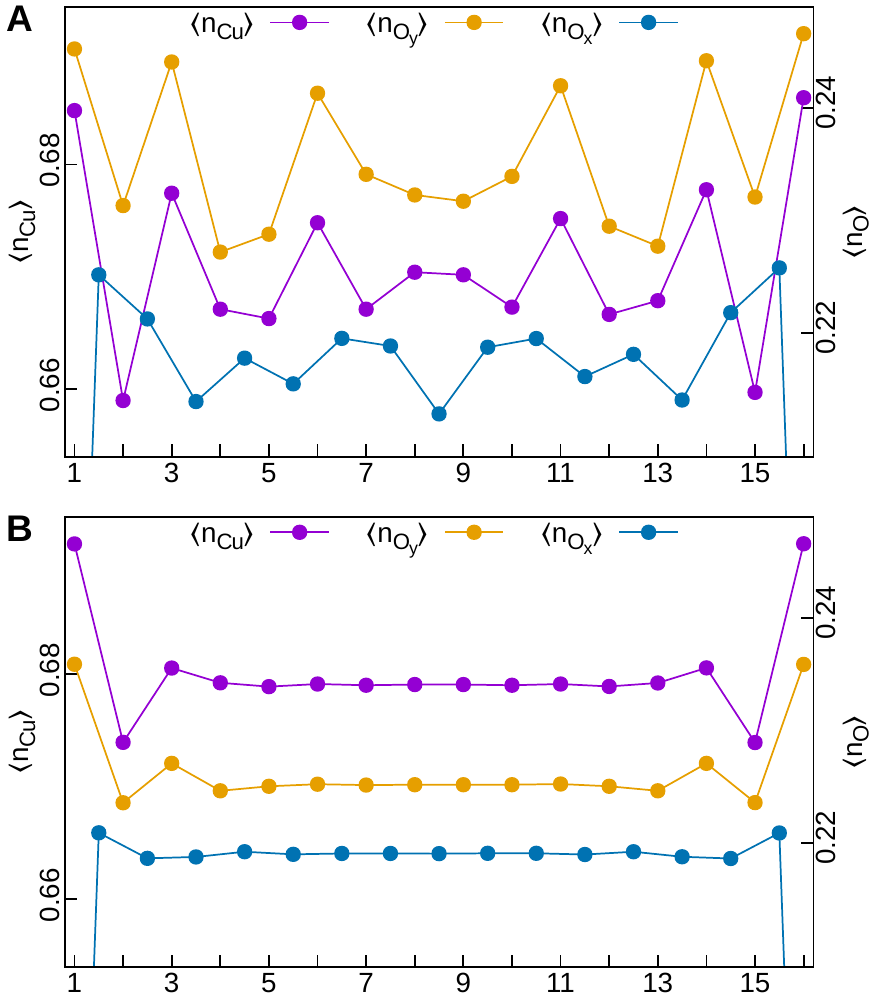}
    \caption{{\bf Comparison of charge modulations in DMRG and DQMC.} ({\bf A}) Orbitally-resolved occupancies as a function of horizontal position, from the DMRG and ({\bf B}) DQMC simulations in Fig.~2 of the main text (where parameters have reduced $U_d$ and $U_p$; see Methods). Errors from random sampling are $O(10^{-5})$ for the DQMC data; hence error bars are invisible on this scale.}
\end{figure}

\newpage
\begin{figure}
    \centering
    \includegraphics{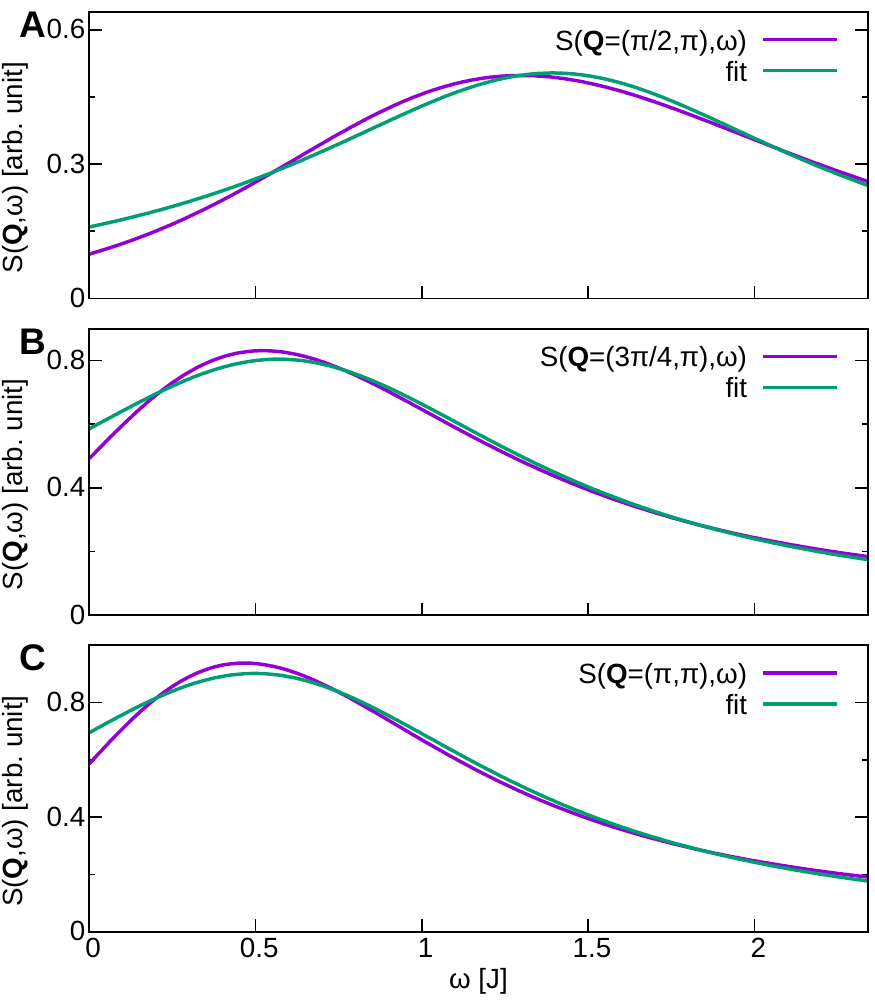}
    \caption{{\bf Fits to EDCs of $\bm{S(Q,\omega)}$.} ({\bf A} to {\bf C}) Examples of fits to EDCs of the $p = 0.125$ data in Fig.~3. A single Lorentzian profile is used.}
\end{figure}

\newpage
\begin{figure}
    \centering
    \includegraphics{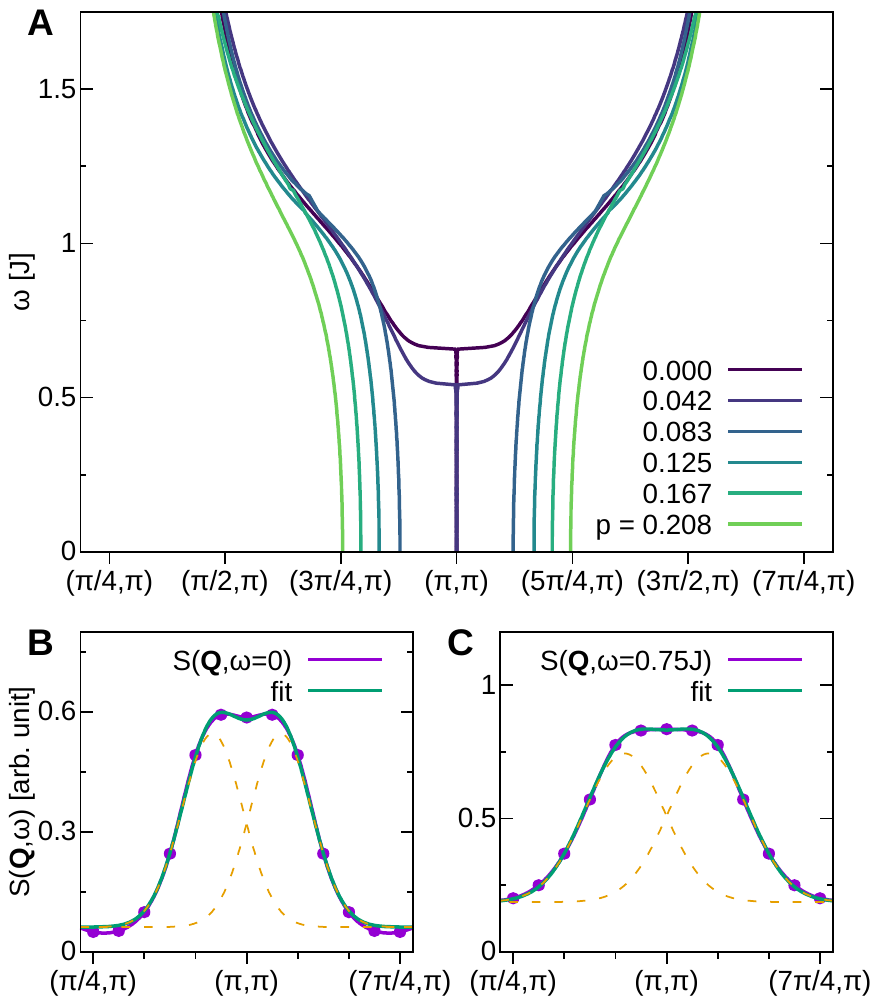}
    \caption{{\bf Fits to MDCs of $\bm{S(Q,\omega)}$.} ({\bf A}) Centers of a double Gaussian fit to MDCs of $S(\bm{Q},\omega)$, for a range of hole dopings. ({\bf B} and {\bf C}) Examples of fits to MDCs of the $p = 0.125$ data. Dots show the raw data from the analytic continuation, with connecting lines of the same color determined via spline interpolation. Dashed lines show the individual Gaussians from the fit.}
\end{figure}

\newpage
\begin{figure}
    \centering
    \includegraphics{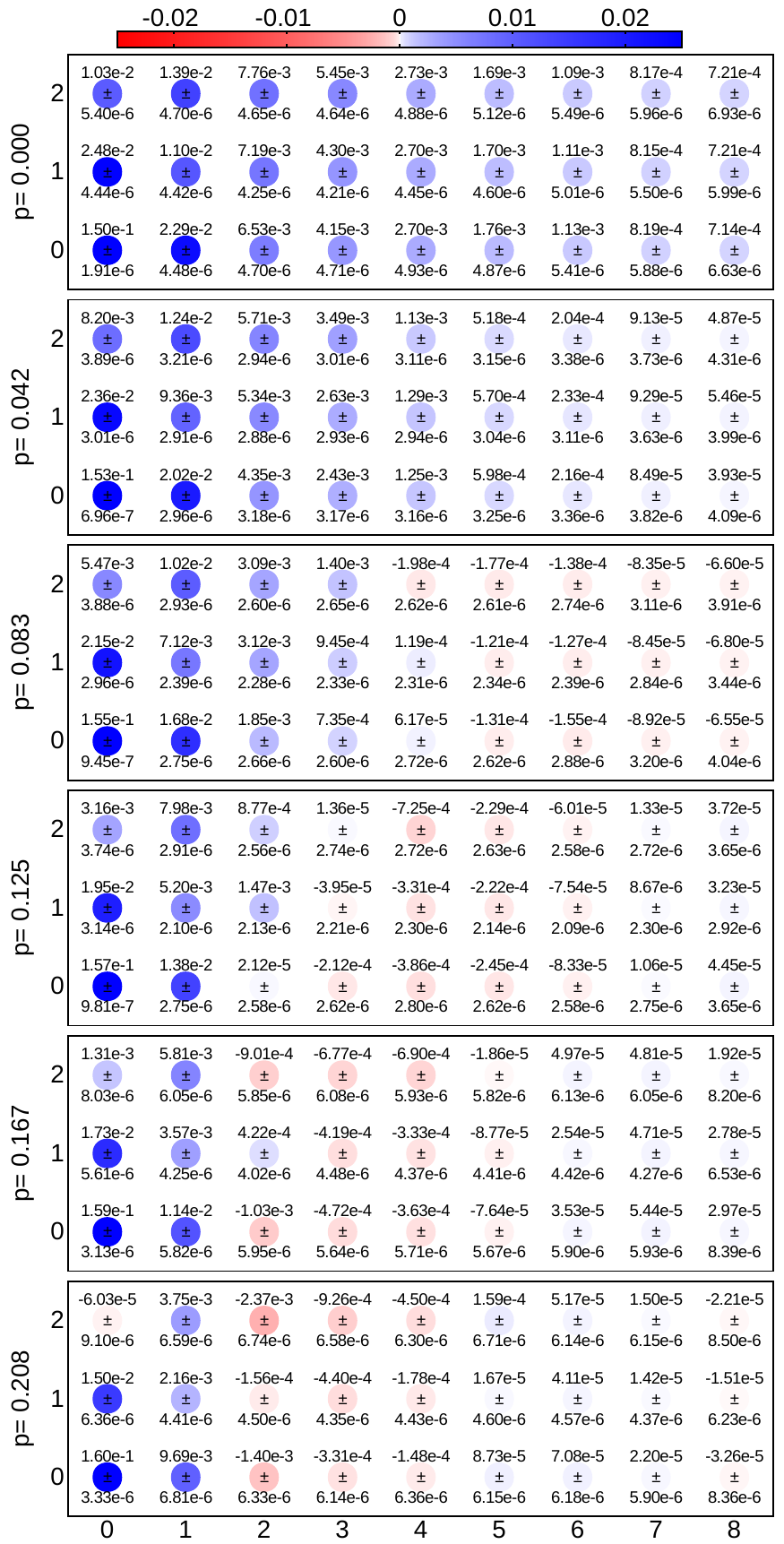}
    \caption{{\bf Values and standard errors of the staggered correlation functions of Fig.~1.} For clarity, only the nonnegative-x, nonnegative-y quadrant is plotted. Data for the remaining quadrants may be inferred by symmetry.}
\end{figure}

\end{document}